\documentstyle[11pt,newpasp,twoside,epsf]{article}
\def\edcomment#1{\iffalse\marginpar{\raggedright\sl#1\/}\else\relax\fi}
\reversemarginpar

\begin{document}
\title{Recycling NSs to ultrashort periods: a statistical analysis of their
evolution in the $\mu  - P$ plane}
\author{Andrea Possenti}
\affil{Dip. Astronomia dell'Universit\`a, via Ranzani 1, 40127 Bologna, Italy}
\author{Monica Colpi}
\affil{Dip. Fisica Mi-Bicocca, P.za della Scienza 3, 20126 Milano, Italy}
\author{Ulrich Geppert}
\affil{AIP, An der Sternwarte 16, 14482 Potsdam, Germany}
\author{Luciano Burderi}
\affil{Oss. Astronomico di Monteporzio, via Frascati 33, 00044 Roma, Italy}
\author{Nichi D'Amico}
\affil{Oss. Astronomico di Bologna, via Ranzani 1, 40127 Bologna, Italy}

\begin{abstract}
We investigate the statistical evolution of magnetic neutron stars 
recycled in Low Mass Binary (LMB) systems, simulating synthetic 
populations. Irrespective to the details of the physical models, 
we find to be significant the fraction of neutron stars spinning 
close to their mass shedding limit relative to the millisecond pulsar 
population. The simulated neutron stars show a tail in their period 
distribution at periods shorter than 1.558 ms, the minimum detected 
so far. Crustal magnetic field decay models predict also the existence 
of massive rapidly spinning neutron stars with very low magnetic moment.
\end{abstract}

\vspace*{-1.0truecm}
\section{Magnetorotational evolution scenario}
\vspace*{-0.2truecm}
Population synthesis models are evolved within a simple 
recycling scenario (as outlined e.g. by Lipunov 1992), 
where the neutron stars (NSs hereon) may experience the phase 
of {\it ejector, accretor or propeller}. The magnetorotational 
evolution includes the general relativistic effects as modelled 
in Burderi et al. (1999). We considered two possibilities 
bracketing uncertainties in the accretion histories of the NSs 
in LMBs: (i) steady accretion for a time $\tau_{RLO};$ 
(ii)  persistent accretion followed by a transient phase during 
which the mass transfer becomes unsteady, mimicking the quenching 
of accretion (Ergma et al. 1998). In particular we account for 
the angular momentum losses by propeller, both for persistent 
and for non-stationary accretion, during the Roche Lobe Overflow phase.

The NSs in our population synthesis models have an initial 
gravitational mass of $1.4~{\rm M_{\odot}}.$ Cook, Shapiro 
\& Teukolsky (1994) showed that all viable equations of state
(EoSs) for nuclear matter allow for recycling to ultra-short periods,
in the case of a nonmagnetic NS. The minimum period below which
the mass-shedding instability limit is encountered, depends however 
on the equation of state. We consider either a soft (FP) and a
stiff (PS) EoS, whose critical periods for mass-shedding are 
$P_{sh}=0.73~$ms (for FP-EoS) and of $P_{sh}=1.40~$msec (for PS-EoS), 
respectively. 

A basic assumption for the present investigation is the crustal origin of the 
NS  magnetic field. Its subsequent evolution is governed by the 
induction equation for ohmic diffusion and advection (see Urpin et al. 1998).
While at the surface the standard boundary condition for a dipolar field 
applies, the inner boundary condition is subject of scientific debates.
Hence we explored two hypotheses, imposing at the crust--core boundary 
either complete field expulsion by the superconducting
core (Boundary condition I, BC I) or advection and freezing (Konar \&
Bhattacharya 1997) in a very highly conducting transition shell (BC II).

In paper I, Possenti et al. (1998) evolved NSs in low mass binaries 
(LMBs) through the five phases of the recycling scenario, and 
explored the evolution of either a core field threaded in the 
proton fluxoids, and a surface field buried by the diamagnetic 
accreting plasma. Here, we start with a population of NSs at the onset 
of the Roche Lobe Overflow (RLO) phase and include the physical evolution 
of a crustal field. The adopted parameters are: 

\vspace*{-0.2truecm}
\begin{table}
\begin{small}
\begin{tabular}{|c||c|c|c|}
{Physical quantity}         & {Distribution} &              {Values }                             &   {Units}
\\
\hline
\hline
 NS period at ${{\rm t_{0}^{RLO}}~^{(a)}}$ &     Flat     &                  1 $~~\to~~$
100                      &   sec
\\
 NS $\mu$ at ${{\rm t_{0}^{RLO}}~^{(a)}}$  &   Gaussian   & Log$<\mu_0>$ =
$~$28.50$~~;~~\sigma$=0.32              &  G~cm$^{3}$
\\
${\dot{m}}$ in RLO phase$~^{(b)}$ &   Gaussian   & Log$<{\dot{m}}>$ =~ --~1.00$~~;~~\sigma$=0.50       & $\rm{{\dot M}_{E}}$
\\
 Minimum accreted mass                 &   One-value  &                       0.01                            & ${\rm {M_{\odot}}}$
\\
 RLO accretion time~$^{(c)}$          &  Flat in Log &
 10$^{6}~~\to~~\tau_{RLO}^{max}~^{(d)}$ &    year
\\
 MSP phase time                        &  Flat in Log &         
 10$^{8}~~\to~~$3$~\times~$10$^{9}$           &    year
\\
\hline
\hline
\end{tabular}
\noindent
$~$
\\
$^{(a)~}$ ${\rm t_{0}^{RLO}}$ = initial time of the RLO phase
\\
$^{(b)~}$ baryonic accretion rate during the RLO phase
\\
$^{(c)~}$ a Maximum accreted Mass of $0.5 M_\odot$ is permitted during the RLO phase
\\
$^{(d)~}$ max duration of the RLO phase; 
typical values: 5$\times$10$^7$ yr - 10$^8$ yr - 5$\times$10$^8$ yr
\\
\hspace*{2.8truecm}
\rule{8cm}{0.02cm}
\\
The values in this Table are derived partly from the observation of LMXBs and partly from the 
tracks in the $\mu - P$ plane calculated by Urpin, Geppert \& Konenkov (1998) 
for a sample of pulsars, under different hypotheses for the binary 
evolution times, for the $\mu$ decay and for the angular momentum 
transfer at the magnetospheric  radius.
\end{small}
\end{table}

\begin{figure}
\plottwo{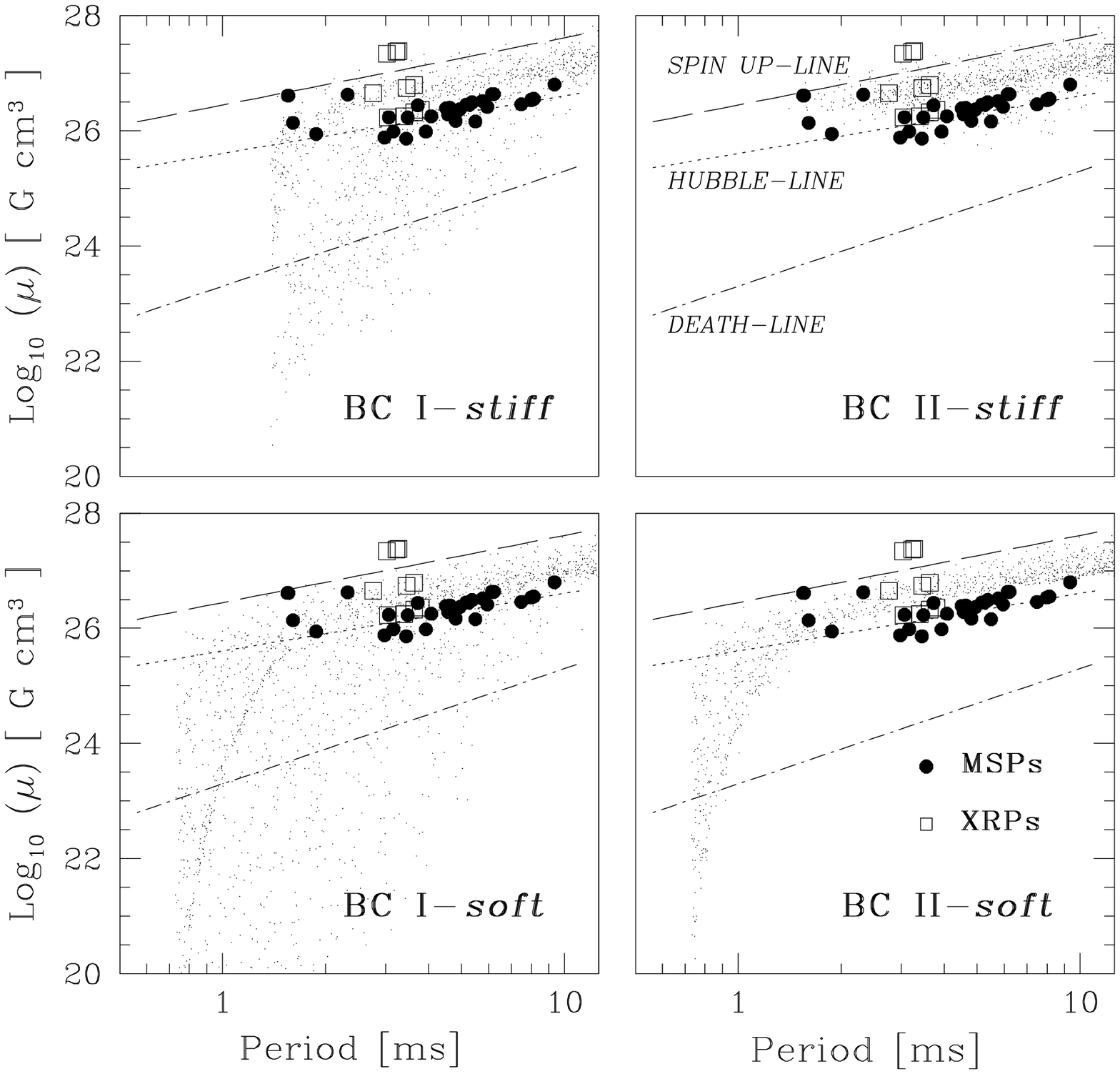}{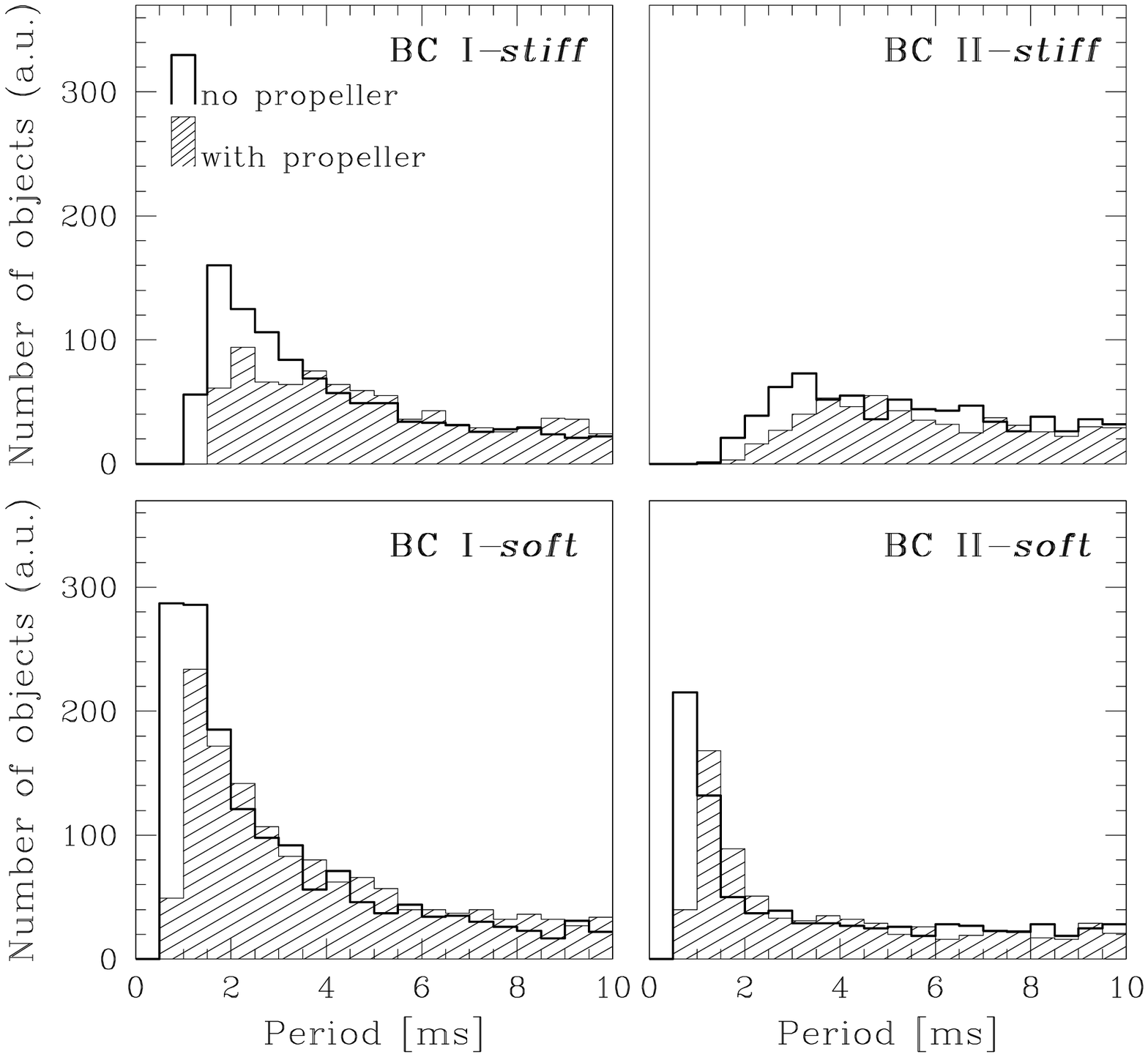}
\caption{{\sf Left:} Statistical properties of the synthesized populations 
in the $\mu-P$ plane, for $<{\dot m}>=0.1$ and 
$\tau_{RLO}^{max}~=~5\times10^8~{\rm yr}$. {\it Full dots} represent the 
sample of detected MSPs, while {\it open squares} represent 
the NSs in LMXBs (White \& Zhang 1997).
{\sf Right:} Calculated distributions (with and without propeller) 
of millisecond NSs as a function of the spin period $P,$ while $\mu$ covers 
the whole range of values.}
\end{figure}

\vspace*{-0.7truecm}
\section{Results and Discussion}
\vspace*{-0.2truecm}
To execute the synthesis calculation we built a Monte Carlo code, 
using typically 3,000 particles. The statistical analysis is carried 
on only those NSs reaching the so-called ``millisecond strip'' 
at the end of recycling, the ones having period $P\leq 10.0$ ms 
and whichever value of the magnetic moment $\mu$. 
In accordance with the values of $P_{min}$ and $\mu_{min}$
(the weakest magnetic moment observed; $\mu_{min}=7.3\times 10^{25}$ 
G cm$^3$), we divided our particles in four groups. Those filling 
the first quadrant in the millisecond strip ($P\geq P_{min}$ and 
$\mu\geq\mu_{min}$) behave as the known MSPs. Also the objects 
belonging to the second quadrant ($P<P_{min}$ and $\mu\geq\mu_{min}$) 
should shines as pulsars (see Burderi \& D'Amico 1997). The 
observability of the objects in the third quadrant ($P<P_{min}$ and 
$\mu < \mu_{min}$) as radio sources represents instead a challenge 
to the modern pulsar surveys. Most of these NSs will   
be above the Chen and Ruderman (1993) ``death-line'', and might have
a bolometric luminosity comparable to that of the known MSPs.  
Thereafter we shortly refer as {\it sub-}MSPs to all the objects 
having $P<P_{min}$ and $\mu$ above the ``death-line''.  
Objects in the fourth quadrant ($P\geq P_{min}$ and $\mu < \mu_{min}$)
are probably radio quiet neutron stars (RQNSs), because they tend to 
be closer to the theoretical ``death-line'', and they are in a period 
range which was already searched with good sensitivity by the radio surveys.

The left panel of Figure 1 illustrates how the simulated NSs fill the
$\mu - P$ plane, while the right panel shows how the EoS affects the shape
of the period distribution. Figure 2 reports on the typical relative 
abundances (left) and gives indication on the role played by the propeller 
(right) in modifying the results.

\begin{figure}
\plottwo{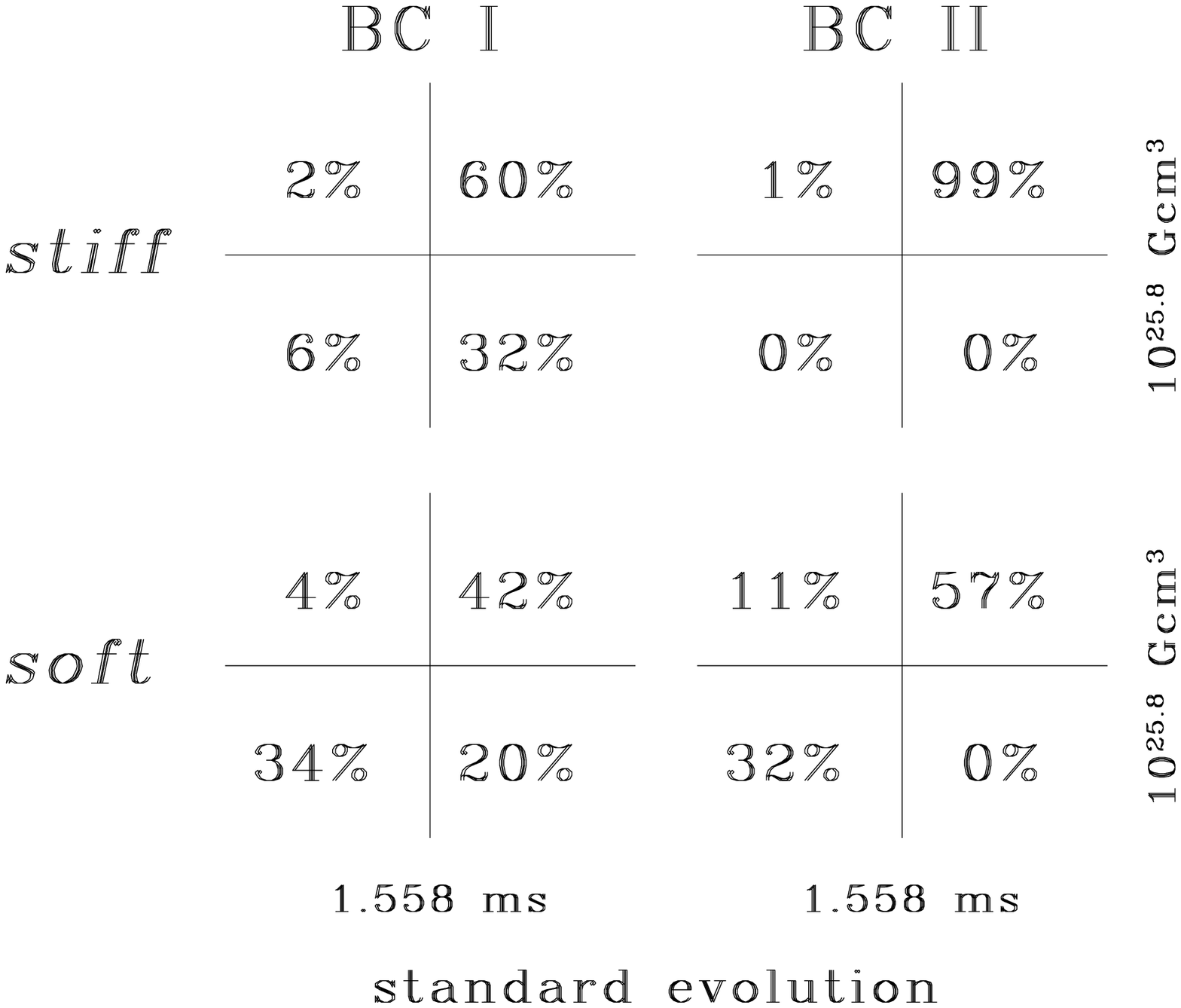}{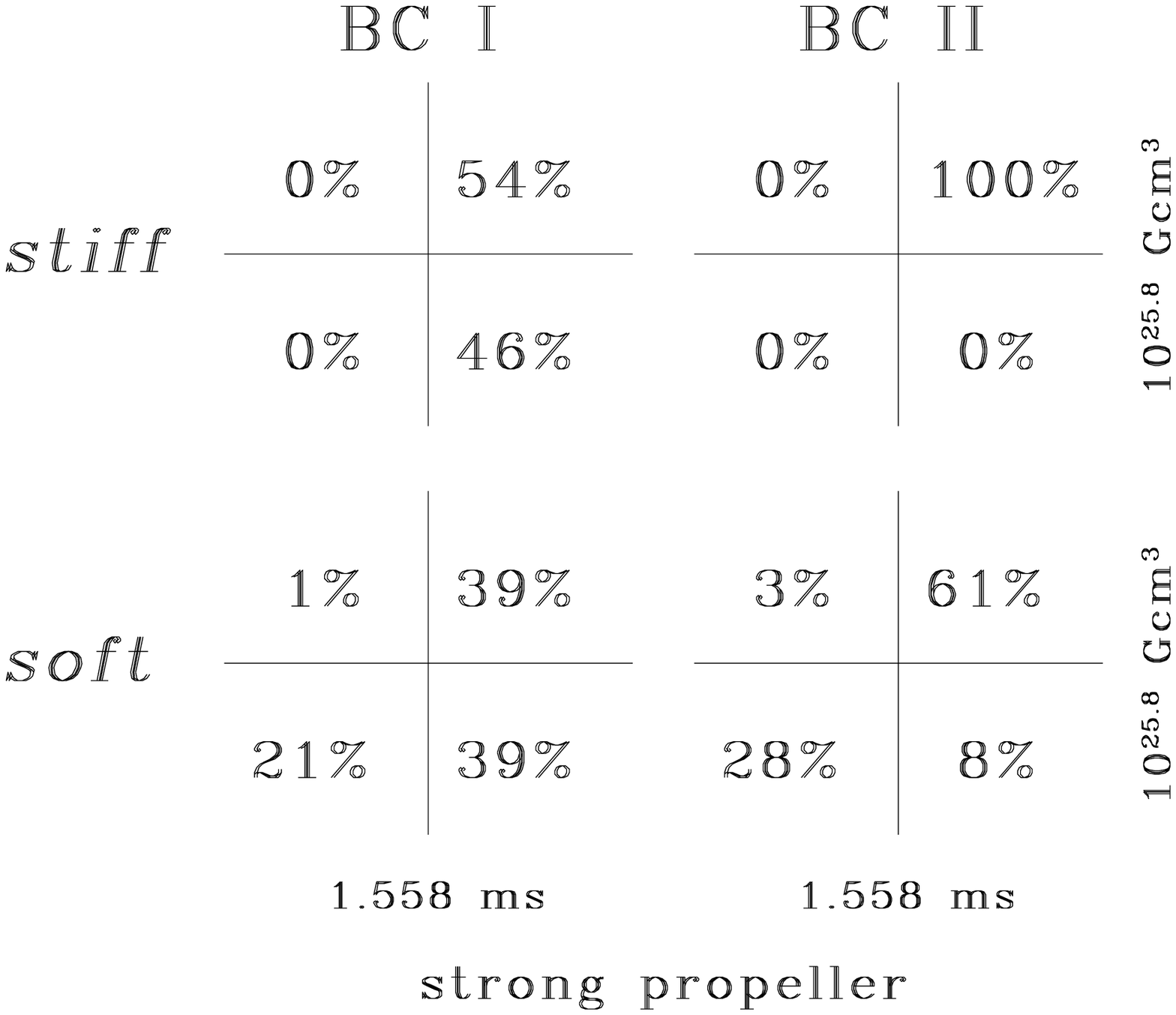}
\caption{Distributions of the synthesized NSs, 
derived normalizing the sample to the total number of stars with 
$P<10~{\rm ms}$. We have divided the $\mu-P$ plane in four regions.
The upper left number in each cross gives the percentage of
objects having $P<P_{min}$ and $\mu>\mu_{min}$. {\sf Left:} Distribution 
for the parameters of Fig. 1-left. {\sf Right:} Distributions when a 
strong propeller is applied, as in Fig. 1-right.}
\end{figure}

\vspace*{-0.6truecm}
\section{Conclusions}
\vspace*{-0.2truecm}

\noindent 
The population synthesis calculation leads to a number of interesting predictions: 

{\sc {\bf A.}}
It shows the presence of a tail in the period distribution of the 
simulated NS populations at periods shorter than $1.558$ ms.

{\sc {\bf B.}}
The fraction of {\it sub-}MSPs estimated over the entire
MSPs population varies between 0 to $\simeq 50$\%. The EoS
and the parameters for the recycling play both an important 
role in determining these percentages. 

{\sc {\bf C.}}  
For a mild-soft EoS and irrespective to the boundary
condition at the crust-core interface for the magnetic field evolution,
the recycling in LMBs gives rise to a NS distribution 
increasing toward short periods and a clear barrier is present 
at the mi\-ni\-mum period for mass-shedding. 

{\sc {\bf D.}}  
For a stiff EoS, the distribution is flatter and displays 
a broad maximum. The maximum is located where most of the spin 
periods of the NSs in LMXBs are found, as indirectly inferred form
the quasi periodic oscillations seen in the timing analysis of 
their X-ray light curves.
  
{\sc {\bf E.}}
If NSs at the end of persistent accretion in a LMB experience a phase
of smooth decline of the accretion rate, the magnetospheric propeller 
at the end of the RLO phase produces a depletion of 
fastly spinning NSs but, at least for the soft EoS, 
it preserves a distribution that peaks at periods $\sim~{\rm 1.5~ms}$.

{\sc {\bf F.}}
The models for the decay of a crustal magnetic field predict the existence
of spun up NSs with very low magnetic moment: their period distribution
is a neat signature for the physics at the crust--core interface.

\noindent
This investigation shows that the detection of {\it sub-}MSPs   
represents a serious challenge to modern radio searches.


\vspace*{-0.4truecm}
\begin{references}
\vspace*{-0.4truecm}
\reference{Burderi, L., \& D'Amico, N. 1997, \apj, 490, 343} 
\reference{Burderi, L., Possenti, A., Colpi, M., Di Salvo, T., D'Amico,
N. 1999, 
\apj, 519, 285}
\reference{Chen, K., \& Ruderman, M. 1993, \apj, 402, 264}
\reference{Cook, G.B., Shapiro S.L., \& Teukolsky, S.A. 1994, 
\apj, 423, L117}
\reference{Ergma, E., Sarna, M.J. \& Antipova, J. 1998, 
\mnras, 300, 352}
\reference{Konar, S., \& Bhattacharya, D., 1997, 
\mnras, 284, 311}
\reference{Lipunov, V.M. 1992, in Astrophysics of neutron stars, 
Springer, Berlin}
\reference{Possenti, A., Colpi, M., D'Amico, N., \& Burderi, L. 1998, 
\apj, 497, L97}
\reference{Urpin, V.A., Geppert U., \& Konenkov, D. 1998, 
\mnras, 295, 907}
\reference{White, N.E., \& Zhang, W. 1997, 
\apj, 490, L87}
\end{references}
\end{document}